\documentclass[%
 reprint,
superscriptaddress,
prl,
 amsmath,amssymb,
 aps
]{revtex4-2}
\usepackage[english]{babel}
\usepackage{amsthm}

\usepackage{graphicx}
\usepackage{dcolumn}
\usepackage{bm}
\usepackage{algpseudocode}
\usepackage{algorithm}
\usepackage{algorithmicx}

\DeclareMathOperator*{\argmin}{argmin}

\newcommand{\s}{{\boldsymbol{s}}}

\newcommand{\x}{{\boldsymbol{x}}}
\newcommand{\w}{{\boldsymbol{w}}}
\newcommand{\1}{{\boldsymbol{1}}}
\newcommand{\vomega}{{\boldsymbol{\omega}}}

\begin{document}

\preprint{APS/123-QED}

\title{Boltzmann sampling with quantum annealers via fast Stein correction}

\author{Ryosuke Shibukawa}
\affiliation{Graduate School of Frontier Sciences, The University of Tokyo, Chiba 277-8568, Japan}
\author{Ryo Tamura}%
\email{tamura.ryo@nims.go.jp}
\affiliation{Graduate School of Frontier Sciences, The University of Tokyo, Chiba 277-8568, Japan}
\affiliation{Center for Basic Research on Materials, National Institute for Materials Science, Ibaraki 305-0044, Japan}
\affiliation{RIKEN Center for Advanced Intelligence Project, Tokyo 103-0027, Japan}
\author{Koji Tsuda}
\email{tsuda@k.u-tokyo.ac.jp}
\affiliation{Graduate School of Frontier Sciences, The University of Tokyo, Chiba 277-8568, Japan}
\affiliation{Center for Basic Research on Materials, National Institute for Materials Science, Ibaraki 305-0044, Japan}
\affiliation{RIKEN Center for Advanced Intelligence Project, Tokyo 103-0027, Japan}

\date{\today}

\begin{abstract}
  Despite the attempts to apply a quantum annealer to Boltzmann sampling,
  it is still impossible to perform accurate sampling at arbitrary temperatures.
  Conventional distribution correction methods such as importance sampling
  and resampling cannot be applied, because the analytical expression of
  sampling distribution is unknown for a quantum annealer. Stein correction (Liu and Lee, 2017)
  can correct
  the samples by weighting without the knowledge of the sampling distribution,
  but the naive implementation requires the solution of a large-scale
  quadratic program, hampering usage in practical problems.
  In this letter, a fast and approximate method based on random feature map and
  exponentiated gradient updates is developed
  to compute the sample weights,
  and used to correct the samples generated by D-Wave quantum annealers.
  In benchmarking problems, it is observed that the residual error of thermal average calculations is reduced significantly.
  If combined with our method, quantum annealers may emerge
  as a viable alternative to long-established Markov chain Monte Carlo methods.   
\end{abstract}

\maketitle



{\it Introduction.}---Boltzmann sampling of the Ising model is central in the studies of critical phenomena~\cite{RevModPhys.58.801,BINDER2001179} and machine learning~\cite{ZHANG20181186,Melko:2019aa}.
Although Markov chain Monte Carlo (MCMC) methods can generate samples according to Boltzmann distribution,
they often fall short for large models due to slow mixing~\cite{doucet2001sequential}. 
To efficiently perform Boltzmann sampling, various improvement techniques such as exchange Monte Carlo and population annealing have been proposed in statistical physics~\cite{hukushima1996exchange,doi:10.1142/S0129183101001912,10.1063/1.1632130}.

As an alternative,
quantum annealers (QAs)~\cite{Johnson:2011aa} have been expected to work as a means to achieve accurate
Boltzmann sampling~\cite{Nelson2022-hb,Raymond2016-je,Marshall2019-fp,Marshall2017-rv,Sandt2023-vu,Vuffray2022-gr}.  
Theoretically, it has been shown that the distribution of quantum annealing samples deviates
from Boltzmann distribution~\cite{Matsuda2009-cs}. 
Nevertheless, scientific discussion is still unsettled about whether QA samples
can be used as samples of a Boltzmann distribution in practical terms.
Recently, Nelson et al. argued that the D-Wave quantum annealer works as an accurate sampler
at certain temperature~\cite{Nelson2022-hb}, but it does not work well at arbitrary temperatures. 
Conventional distribution correction techniques such as
importance sampling and resampling~\cite{doucet2001sequential} cannot be applied,
because the sampling distribution of a quantum annealer has not been analytically described so far.

Liu and Lee proposed a ``black-box'' distribution correction method based on Stein statistics,
where the analytical form of the original distribution is not needed~\cite{Liu2017-nn}.
The original samples are assigned the weights to fit to the target distribution via quadratic programming.
In the original paper, the theoretical properties are largely unsolved,
but Hodgkinson et al. showed the convergence of Stein correction for samples generated by a Markov
chain~\cite{Hodgkinson2020-cr}.
This letter investigates how well this method, called Stein correction,
works in the distribution correction of QA samples to a Boltzmann distribution with a given temperature.   
First, we develop a fast approximate algorithm of Stein correction,
because $O(n^3)$ computational cost of quadratic programming for the number of samples, $n$, is prohibitive for
a large number of samples.  
In benchmarking studies, we observed that the estimation error of internal energy,
magnetic susceptibility, and Binder cumulant of some Ising models decreased in a large extent by Stein correction.
This result implies that Stein correction is useful for improving sample quality
for applications such as critical phenomena and machine learning.
It can also be applied to general quantum computers including NISQ,
where distributional error is unavoidable due to environmental noise~\cite{pelofske2021sampling}.


{\it Fast Stein correction.}---Denote by $p(\x), q(\x)$ two distributions in $\x \in \{x_i = -1,1\}^d$.
Let $\neg_i$ denote the sign flip of the $i$-th variable.  
Kernelized Stein discrepancy~\cite{Yang2018-pa} quantifies
the difference between the two distributions as
\begin{equation} \label{eq:ksd}
S(p,q) = {\mathbb E}_{\x,\x^\prime \sim q} [k_p(\x,\x^\prime)],
\end{equation}  
where $k_p(\x,\x^\prime)$ is called Stein kernel that depends on $p(\x)$
and the base kernel
\begin{equation} \label{eq:base}
k(\x, \x^\prime) = \exp \left(-\frac{\sum_{i=1}^d \mathbb{I}\{x_i \neq x^\prime_i\}}{d} \right).
\end{equation}
The function of $\mathbb{I}\{x_i \neq x^\prime_i\}$ shows 1 for $x_i \neq x^\prime_i$ and 0 for the others, respectively.
Let us define the difference operator as
\begin{equation}
\nabla_{\bm{x}} f(\bm x) = (f(\x)-f(\neg_1 \x),\ldots,f(\x) - f(\neg_d \x)). 
\end{equation}
In addition, the score function $\s_p(\x) \in \mathbb{R}^d$ is defined as  
\[
[\s_p(\x)]_i  = 1 - p(\neg_i \x)/p(\x), \ \ (i=1,\ldots,d).
\]
The Stein kernel is then defined as
\begin{align}
k_p(\x_i, \x_j) &= \s_p(\x_i)^\top k(\x_i, \x_j) \s_p (\x_j) \nonumber\\
           &- \s_p(\x_i)^\top \nabla_{\x_j} k(\x_i, \x_j) \nonumber\\
           &- \s_p(\x_j)^\top \nabla_{\x_i} k(\x_i, \x_j)\nonumber\\
           &+ \text{Tr} ( \nabla_{\x_i, \x_j}k(\x_i, \x_j) ).
\end{align}
Notably, $k_p(\x_i,\x_j)$ depends on $p(\x)$ only through the score functions.
Therefore, when $p(\x)$ is the Boltzmann distribution,
the Stein kernel does not depend on the normalization constant.
The Stein discrepancy defined by Eq.~(\ref{eq:ksd}) is always nonnegative and zero
if $p(\x)$ and $q(\x)$ are identical~\cite{Hodgkinson2020-cr}.
Given the $n$ samples from $q(\x)$, $\x_1,\ldots,\x_n$,
the discrepancy is approximated as 
\begin{equation}
{\hat S}(p,q) = \sum_{i,j=1}^n w_i w_j k_p(\x_i,\x_j), 
\end{equation}
where $w_1, \ldots, w_n$ are the weight for each sample.
In Stein correction~\cite{Liu2017-nn}, the weights are adjusted such that
the discrepancy is minimized,
\begin{equation}\label{eq:quad}
    \hat{\w} = \argmin_{\w}\left\{\w ^ \top K_p \w \text{  s.t. } w_i \ge 0, \sum_{i=1}^{n} w_i = 1 \right\},
\end{equation}
where $K_p$ is the $n \times n$ matrix where elements are $k_p(\x_i, \x_j)$,
which is called Stein kernel matrix.

A naive implementation of Stein correction
requires $O(n^2)$ space and $O(n^3)$ time.
We reduce the complexity by introducing random feature map~\cite{Rahimi2007-ay}
and exponentiated gradient descent~\cite{Kivinen1997-va}.
Using the random feature map, the base kernel defined by Eq.~(\ref{eq:base}) is
approximated as the inner product $k(\x,\x^\prime) \approx \phi(\x)^\top \phi(\x^\prime)$
where the feature map $\phi(\x): \{-1,1\}^d \rightarrow \mathbb{R}^\ell$ is computed as follows. 
Let us draw $\ell$ samples $\vomega_1,\ldots,\vomega_\ell$ from
\begin{equation}
h(\vomega) = \prod_{i=1}^d \frac{1}{\pi(1+\omega_i^2)},
\end{equation}
where $\vomega$ is the $d$-dimensional vector with each component $\omega_i$.
Also, $b_1,\ldots,b_\ell$ are sampled from the uniform distribution over $[0,2\pi]$. 
Then, the feature map is defined as
\begin{equation}
\phi(\x) = \left( z_{\vomega_1, b_1}\left(\frac{\x+{\bf1}}{2d}\right), \ldots, z_{\vomega_\ell,b_\ell}\left(\frac{\x+{\bf1}}{2d} \right) \right)^\top,
\end{equation}
where $z_{\vomega,b}(\x) = \sqrt{\frac{2}{\ell}} \cos(\vomega^\top \x + b)$
and ${\bf 1}$ is the $d$-dimensional vector of $(1,\ldots,1)$.
Since the Stein kernel is a linear function of the base kernel,
it can also be approximated as
$k_p(\x,\x^\prime) \approx \phi_p(\x)^\top \phi_p(\x)$,
where $\phi_p(\x)$ is the concatenation of the following vectors:
\begin{equation}
\theta_k(\x) = \frac{p(\neg_k \x)}{p(\x)} \phi(\x) - \phi(\neg_k \x), \ \ \ (k=1,\ldots,d). 
\end{equation}
Using the random feature map, the optimization problem defined by Eq.~(\ref{eq:quad}) is rewritten as follows:
\begin{equation}
    \hat{\w} = \argmin_{\w}\left\{ f(\w) \text{  s.t. } w_i \ge 0, \sum_{i=1}^{n} w_i = 1 \right\},
\end{equation}
where $f(\w) = \| \sum_{i=1}^n w_i \phi_p(\x_i) \|^2$.
This is a convex optimization problem with nonnegativity
and normalization constraints.
When the standard gradient descent algorithm is applied,
the constraints are violated every time the parameters are updated.
In this case, exponentiated gradient descent is known to work well~\cite{Kivinen1997-va},
because constraint violation never happens. The update is described as
\begin{equation}
w_{t+1,i} = \frac{w_{t,i} \exp(-\eta [\nabla f(\w_t)]_i)}{Z_t},
\end{equation}
where
\begin{equation}
Z_t = \sum_{i=1}^n w_{t,i} \exp(-\eta [\nabla f(\w_t)]_i),
\end{equation}
and $\eta$ is the learning rate.
The modification shown above reduces the space requirement to $O(n \ell)$.
Each update takes only $O(n)$ time, enabling us to deal with a large number of samples.
The implementation of a fast Stein correction can be found on GitHub \url{https://github.com/tsudalab/fast-stein-correction}.

\begin{figure*}
  \begin{center}
  \includegraphics[width=0.99\textwidth]{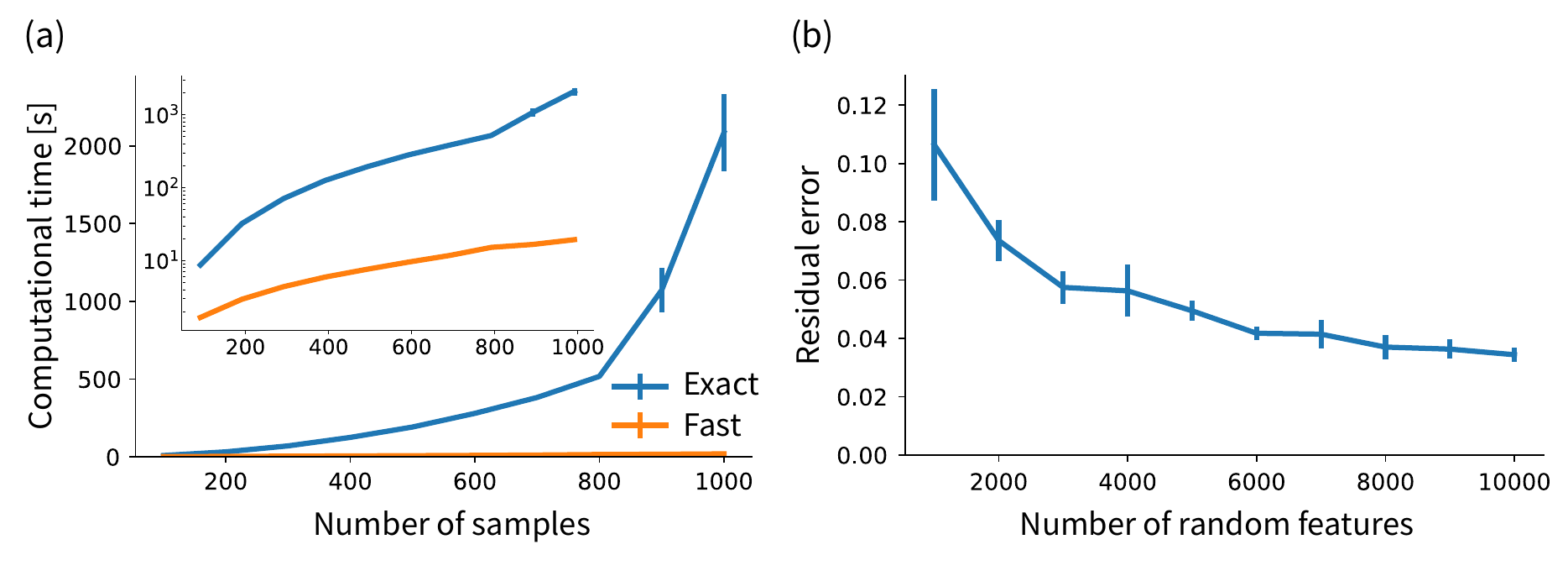}
  \end{center}
  \caption{Results for Stein correction on  GSD\_8. (a) Computational time of exact and fast Stein correction depending on the number of samples $n$ when $\ell = 5,000$. The inset is the log scale figure. 
  (b) Residual error of the Stein kernel matrix $K_p$ against the number of random features $\ell$ when $n=1,000$.
  Five independent runs are performed, and the mean and standard deviation are plotted as lines and error bars, respectively.}
  \label{fig:res0}
\end{figure*}


{\it Boltzmann sampling.}---We are engaged in sampling from the Boltzmann distribution,
$p(\x) \sim \exp[- \beta H_{{\rm Ising}}(\x)]$, $\x \in \{-1,1\}^d$,  
where the Hamiltonian is described as
\begin{equation}
H_{{\rm Ising}}(\x) = - \sum_{i,j \in E} J_{ij} x_i x_j - \sum_{i \in V} h_i x_i.
\end{equation}
$\beta$ denotes the inverse temperature, $V \subseteq [1,d]$
and $E \subseteq [1,d] \times [1,d]$.  
Here, we assume that the parameters $J_{ij}$ and $h_i$ are in the range of $-1$ and $1$.
The thermal average of observable $\mathcal{O} (\x)$ is defined by 
\begin{eqnarray}
\langle \mathcal{O} (\x) \rangle_\beta = \frac{{\rm Tr} \mathcal{O} (\x) \exp [- \beta H_{\rm{Ising}}(\x)]}{{\rm Tr}\exp [- \beta H_{\rm{Ising}}(\x)]}.
\end{eqnarray}
Since the trace calculation is impossible for large models,
this trace is approximately replaced to the average of some samples.
go
When $\beta H_{{\rm Ising}}(\x)$ is solved by QA with short annealing time,
the distribution of samples is more wide spread around the ground state.
If we consider that this distribution is Boltzmann distribution, the thermal average is calculated as
\begin{eqnarray}
\langle \mathcal{O} (\x) \rangle_\beta^{\rm QA} = \frac{1}{n}\sum_{i=1}^n \mathcal{O} (\x_i), \label{TA_QA}
\end{eqnarray}
where $\x_1,\ldots,\x_n$ are the samples by QA.
But, the distribution by QA deviates from the Boltzmann distribution,
and the correction is needed.
By performing the Stein correction where $p(\x)$ and $q(\x)$ are the Boltzmann distribution with $\beta$ and the distribution by QA, respectively, weights $\hat{w}_i$ for each sample are evaluated.
Using the weights, the thermal average is approximately obtained by
\begin{eqnarray}
\langle \mathcal{O} (\x) \rangle_\beta^{\rm SC} = \sum_{i=1}^n \hat{w}_i \mathcal{O} (\x_i). \label{TA_SC}
\end{eqnarray}

{\it Results.}---In our experiments, we employ 16-bit Ising Hamiltonians
proposed by Nelson et al.\cite{Nelson2022-hb} for benchmarking.
They are called GSD\_X and the number X indicates
the number of degenerated ground states.
The Hamiltonians are designed to fit the Chimera topology of D-Wave 2000Q systems,
but here the samples are generated by Advantage System 6.2,
because 2000Q is already out of service in their cloud platform.
Hamiltonians are embedded in the Pegasus topology of Advantage
using minorminer python package~\cite{minorminer}.

First, we evaluate the efficiency and approximation error of
fast Stein correction using GSD\_8.
The residual error of the Stein kernel matrix $K_p$
is defined as $\| K_p - K^*_p \| / \| K^*_p \|$,
where $K^*_p$ is the exact matrix.
Throughout the letter,
the learning rate is $\eta=10^{-5}$ and the number of updates is $3,000$. 
An exact correction was performed by cvxopt~\cite{cvxopt}. 
Figure~\ref{fig:res0}a shows the computation time of Stein correction depending on the number of samples when $\ell =5,000$.
The fast correction was orders of magnitude faster than the exact one against the number of samples.
The residual error by the random feature map decreased
as the number of features was increased (see Fig.~\ref{fig:res0}b),
showing that the more random features are preferred for accurate approximation.
In the following analysis, calculations are performed with $\ell=5,000$, indicating a sufficiently better residual error and taking computational time into account.

\begin{figure*}
  \begin{center}
  \includegraphics[width=0.84\textwidth]{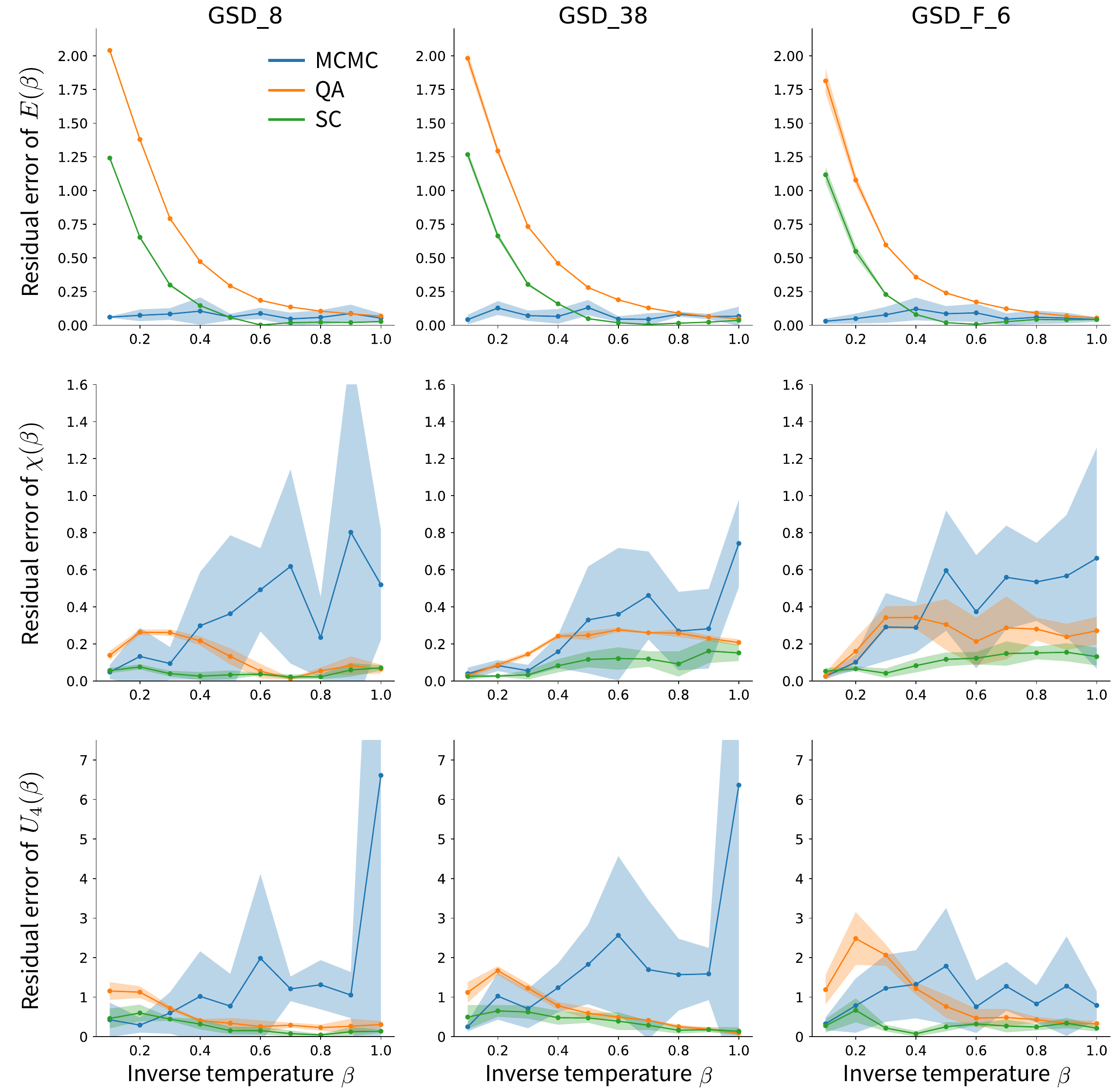}
  \end{center}
  \caption{Residual errors of internal energy $E(\beta)$, magnetic susceptibility $\chi(\beta)$, and Binder cumulant $U_4(\beta)$.
  The thermal averages are calculated by MCMC, a naive average by quantum annealer (QA), and Stein correction (SC), respectively.
  The number of samples is fixed as $n=10,000$, and the results depending on $n$ are shown in Fig.~S1.
  Five independent runs are performed, and the mean and standard deviation are plotted as lines and shaded areas, respectively.}
\label{fig:res1}
\end{figure*}

Next, we observed how fast Stein correction improves the accuracy of estimating
observables on GSD\_8, GSD\_38, and GSD\_F\_6.
For GSD\_F\_6, the finite fields $h_i$ are imposed.
The calculated thermal averages of observable are internal energy $E(\beta)$, magnetic susceptibility $\chi(\beta)$, and Binder cumulant $U_4(\beta)$ defined by
\begin{align}
E(\beta) &= \langle H_{\rm Ising} (\x) \rangle_\beta, \\ 
\chi(\beta) &= \beta \langle (\textstyle{\sum_{i=1}^d x_i)^2} \rangle_\beta, \\ 
U_4(\beta) &= 1 - \textstyle{\langle (\sum_{i=1}^d x_i)^4 \rangle_\beta / 3 \langle (\sum_{i=1}^d x_i)^2 \rangle_\beta^2},
\end{align}
where $d$ is the system size.
For each observable, we compute the residual error as $\| y-y^* \|/y^*$,
where $y$ is the thermal average of observables calculated by Eqs.~(\ref{TA_QA}) and (\ref{TA_SC}) and $y^*$ is the exact value computed
via brute-force enumeration.
For each $\beta$, 10,000 samples are generated by a D-Wave quantum annealer.
In addition, the Metropolis method, a basic MCMC method,
was also applied. Here, the first 8,000 samples were used for burn-in
and the remaining ones for estimation.  

The results at annealing time $5\mu s$ over a range of inverse temperatures
are shown in Fig.~\ref{fig:res1}. 
For each case, 5 independent runs are conducted and the mean values of residual error are evaluated.
The accuracy of fast Stein correction outperformed the original quantum annealing
samples in all cases, showing the effectiveness of our approach.
Furthermore, the error by fast Stein correction was
consistently smaller than that by MCMC.
These results show that fast Stein correction has the potential to expand
the applicablity range of quantum annealers significantly
and may replace MCMC in diverse tasks of discrete sampling.

{\it Conclusion.}---In summary, we demonstrated that fast Stein correction is a helpful companion of
quantum annealers and fundamentally enhance their usability.
The advantage of quantum samples is that they are not locally concentrated,
whereas MCMC samples have difficulty to cover the whole space.
Although Stein correction cannot bring the distributional error to zero,
it would be particularly useful to sample from highly constrained spaces~\cite{aoki2012markov},
where global mixing by MCMC is extremely hard. 
Our future work involves the application of our method
to machine learning and statistical physics and
other highly scalable Ising machines
such as coherent Ising machines~\cite{Inagaki:2016aa}
and GPU-based algorithms~\cite{mao2023chemical}.

\begin{acknowledgments}
RS thanks participants at AQC2023 in Albuquerque, New Mexico for fruitful discussions.
This work is supported by AIP Kasoku JPMJCR21U2, JST CREST
JPMJCR21O2, JST ERATO JPMJER1903, KAKENHI 19H05819 and MEXT JPMXP1122712807.
\end{acknowledgments}





\begin{thebibliography}{27}%
\makeatletter
\providecommand \@ifxundefined [1]{%
 \@ifx{#1\undefined}
}%
\providecommand \@ifnum [1]{%
 \ifnum #1\expandafter \@firstoftwo
 \else \expandafter \@secondoftwo
 \fi
}%
\providecommand \@ifx [1]{%
 \ifx #1\expandafter \@firstoftwo
 \else \expandafter \@secondoftwo
 \fi
}%
\providecommand \natexlab [1]{#1}%
\providecommand \enquote  [1]{``#1''}%
\providecommand \bibnamefont  [1]{#1}%
\providecommand \bibfnamefont [1]{#1}%
\providecommand \citenamefont [1]{#1}%
\providecommand \href@noop [0]{\@secondoftwo}%
\providecommand \href [0]{\begingroup \@sanitize@url \@href}%
\providecommand \@href[1]{\@@startlink{#1}\@@href}%
\providecommand \@@href[1]{\endgroup#1\@@endlink}%
\providecommand \@sanitize@url [0]{\catcode `\\12\catcode `\$12\catcode
  `\&12\catcode `\#12\catcode `\^12\catcode `\_12\catcode `\%12\relax}%
\providecommand \@@startlink[1]{}%
\providecommand \@@endlink[0]{}%
\providecommand \url  [0]{\begingroup\@sanitize@url \@url }%
\providecommand \@url [1]{\endgroup\@href {#1}{\urlprefix }}%
\providecommand \urlprefix  [0]{URL }%
\providecommand \Eprint [0]{\href }%
\providecommand \doibase [0]{http://dx.doi.org/}%
\providecommand \selectlanguage [0]{\@gobble}%
\providecommand \bibinfo  [0]{\@secondoftwo}%
\providecommand \bibfield  [0]{\@secondoftwo}%
\providecommand \translation [1]{[#1]}%
\providecommand \BibitemOpen [0]{}%
\providecommand \bibitemStop [0]{}%
\providecommand \bibitemNoStop [0]{.\EOS\space}%
\providecommand \EOS [0]{\spacefactor3000\relax}%
\providecommand \BibitemShut  [1]{\csname bibitem#1\endcsname}%
\let\auto@bib@innerbib\@empty
\bibitem [{\citenamefont {Binder}\ and\ \citenamefont
  {Young}(1986)}]{RevModPhys.58.801}%
  \BibitemOpen
  \bibfield  {author} {\bibinfo {author} {\bibfnamefont {K.}~\bibnamefont
  {Binder}}\ and\ \bibinfo {author} {\bibfnamefont {A.~P.}\ \bibnamefont
  {Young}},\ }\href {\doibase 10.1103/RevModPhys.58.801} {\bibfield  {journal}
  {\bibinfo  {journal} {Rev. Mod. Phys.}\ }\textbf {\bibinfo {volume} {58}},\
  \bibinfo {pages} {801} (\bibinfo {year} {1986})}\BibitemShut {NoStop}%
\bibitem [{\citenamefont {Binder}\ and\ \citenamefont
  {Luijten}(2001)}]{BINDER2001179}%
  \BibitemOpen
  \bibfield  {author} {\bibinfo {author} {\bibfnamefont {K.}~\bibnamefont
  {Binder}}\ and\ \bibinfo {author} {\bibfnamefont {E.}~\bibnamefont
  {Luijten}},\ }\href {\doibase https://doi.org/10.1016/S0370-1573(00)00127-7}
  {\bibfield  {journal} {\bibinfo  {journal} {Phys. Rep.}\ }\textbf {\bibinfo
  {volume} {344}},\ \bibinfo {pages} {179} (\bibinfo {year}
  {2001})}\BibitemShut {NoStop}%
\bibitem [{\citenamefont {Zhang}\ \emph {et~al.}(2018)\citenamefont {Zhang},
  \citenamefont {Ding}, \citenamefont {Zhang},\ and\ \citenamefont
  {Xue}}]{ZHANG20181186}%
  \BibitemOpen
  \bibfield  {author} {\bibinfo {author} {\bibfnamefont {N.}~\bibnamefont
  {Zhang}}, \bibinfo {author} {\bibfnamefont {S.}~\bibnamefont {Ding}},
  \bibinfo {author} {\bibfnamefont {J.}~\bibnamefont {Zhang}}, \ and\ \bibinfo
  {author} {\bibfnamefont {Y.}~\bibnamefont {Xue}},\ }\href {\doibase
  https://doi.org/10.1016/j.neucom.2017.09.065} {\bibfield  {journal} {\bibinfo
   {journal} {Neurocomputing}\ }\textbf {\bibinfo {volume} {275}},\ \bibinfo
  {pages} {1186} (\bibinfo {year} {2018})}\BibitemShut {NoStop}%
\bibitem [{\citenamefont {Melko}\ \emph {et~al.}(2019)\citenamefont {Melko},
  \citenamefont {Carleo}, \citenamefont {Carrasquilla},\ and\ \citenamefont
  {Cirac}}]{Melko:2019aa}%
  \BibitemOpen
  \bibfield  {author} {\bibinfo {author} {\bibfnamefont {R.~G.}\ \bibnamefont
  {Melko}}, \bibinfo {author} {\bibfnamefont {G.}~\bibnamefont {Carleo}},
  \bibinfo {author} {\bibfnamefont {J.}~\bibnamefont {Carrasquilla}}, \ and\
  \bibinfo {author} {\bibfnamefont {J.~I.}\ \bibnamefont {Cirac}},\ }\href
  {\doibase 10.1038/s41567-019-0545-1} {\bibfield  {journal} {\bibinfo
  {journal} {Nat. Phys.}\ }\textbf {\bibinfo {volume} {15}},\ \bibinfo {pages}
  {887} (\bibinfo {year} {2019})}\BibitemShut {NoStop}%
\bibitem [{\citenamefont {Doucet}\ \emph {et~al.}(2001)\citenamefont {Doucet},
  \citenamefont {De~Freitas}, \citenamefont {Gordon} \emph
  {et~al.}}]{doucet2001sequential}%
  \BibitemOpen
  \bibfield  {author} {\bibinfo {author} {\bibfnamefont {A.}~\bibnamefont
  {Doucet}}, \bibinfo {author} {\bibfnamefont {N.}~\bibnamefont {De~Freitas}},
  \bibinfo {author} {\bibfnamefont {N.~J.}\ \bibnamefont {Gordon}},  \emph
  {et~al.},\ }\href@noop {} {\emph {\bibinfo {title} {Sequential Monte Carlo
  methods in practice}}}\ (\bibinfo  {publisher} {Springer},\ \bibinfo {year}
  {2001})\BibitemShut {NoStop}%
\bibitem [{\citenamefont {Hukushima}\ and\ \citenamefont
  {Nemoto}(1996)}]{hukushima1996exchange}%
  \BibitemOpen
  \bibfield  {author} {\bibinfo {author} {\bibfnamefont {K.}~\bibnamefont
  {Hukushima}}\ and\ \bibinfo {author} {\bibfnamefont {K.}~\bibnamefont
  {Nemoto}},\ }\href@noop {} {\bibfield  {journal} {\bibinfo  {journal} {J.
  Phys. Soc. Japan}\ }\textbf {\bibinfo {volume} {65}},\ \bibinfo {pages}
  {1604} (\bibinfo {year} {1996})}\BibitemShut {NoStop}%
\bibitem [{\citenamefont {Iba}(2001)}]{doi:10.1142/S0129183101001912}%
  \BibitemOpen
  \bibfield  {author} {\bibinfo {author} {\bibfnamefont {Y.}~\bibnamefont
  {Iba}},\ }\href {\doibase 10.1142/S0129183101001912} {\bibfield  {journal}
  {\bibinfo  {journal} {Int. J. Mod. Phys. C}\ }\textbf {\bibinfo {volume}
  {12}},\ \bibinfo {pages} {623} (\bibinfo {year} {2001})}\BibitemShut
  {NoStop}%
\bibitem [{\citenamefont {Hukushima}\ and\ \citenamefont
  {Iba}(2003)}]{10.1063/1.1632130}%
  \BibitemOpen
  \bibfield  {author} {\bibinfo {author} {\bibfnamefont {K.}~\bibnamefont
  {Hukushima}}\ and\ \bibinfo {author} {\bibfnamefont {Y.}~\bibnamefont
  {Iba}},\ }\href {\doibase 10.1063/1.1632130} {\bibfield  {journal} {\bibinfo
  {journal} {AIP Conference Proceedings}\ }\textbf {\bibinfo {volume} {690}},\
  \bibinfo {pages} {200} (\bibinfo {year} {2003})}\BibitemShut {NoStop}%
\bibitem [{\citenamefont {Johnson}\ \emph {et~al.}(2011)\citenamefont
  {Johnson}, \citenamefont {Amin}, \citenamefont {Gildert}, \citenamefont
  {Lanting}, \citenamefont {Hamze}, \citenamefont {Dickson}, \citenamefont
  {Harris}, \citenamefont {Berkley}, \citenamefont {Johansson}, \citenamefont
  {Bunyk}, \citenamefont {Chapple}, \citenamefont {Enderud}, \citenamefont
  {Hilton}, \citenamefont {Karimi}, \citenamefont {Ladizinsky}, \citenamefont
  {Ladizinsky}, \citenamefont {Oh}, \citenamefont {Perminov}, \citenamefont
  {Rich}, \citenamefont {Thom}, \citenamefont {Tolkacheva}, \citenamefont
  {Truncik}, \citenamefont {Uchaikin}, \citenamefont {Wang}, \citenamefont
  {Wilson},\ and\ \citenamefont {Rose}}]{Johnson:2011aa}%
  \BibitemOpen
  \bibfield  {author} {\bibinfo {author} {\bibfnamefont {M.~W.}\ \bibnamefont
  {Johnson}}, \bibinfo {author} {\bibfnamefont {M.~H.~S.}\ \bibnamefont
  {Amin}}, \bibinfo {author} {\bibfnamefont {S.}~\bibnamefont {Gildert}},
  \bibinfo {author} {\bibfnamefont {T.}~\bibnamefont {Lanting}}, \bibinfo
  {author} {\bibfnamefont {F.}~\bibnamefont {Hamze}}, \bibinfo {author}
  {\bibfnamefont {N.}~\bibnamefont {Dickson}}, \bibinfo {author} {\bibfnamefont
  {R.}~\bibnamefont {Harris}}, \bibinfo {author} {\bibfnamefont {A.~J.}\
  \bibnamefont {Berkley}}, \bibinfo {author} {\bibfnamefont {J.}~\bibnamefont
  {Johansson}}, \bibinfo {author} {\bibfnamefont {P.}~\bibnamefont {Bunyk}},
  \bibinfo {author} {\bibfnamefont {E.~M.}\ \bibnamefont {Chapple}}, \bibinfo
  {author} {\bibfnamefont {C.}~\bibnamefont {Enderud}}, \bibinfo {author}
  {\bibfnamefont {J.~P.}\ \bibnamefont {Hilton}}, \bibinfo {author}
  {\bibfnamefont {K.}~\bibnamefont {Karimi}}, \bibinfo {author} {\bibfnamefont
  {E.}~\bibnamefont {Ladizinsky}}, \bibinfo {author} {\bibfnamefont
  {N.}~\bibnamefont {Ladizinsky}}, \bibinfo {author} {\bibfnamefont
  {T.}~\bibnamefont {Oh}}, \bibinfo {author} {\bibfnamefont {I.}~\bibnamefont
  {Perminov}}, \bibinfo {author} {\bibfnamefont {C.}~\bibnamefont {Rich}},
  \bibinfo {author} {\bibfnamefont {M.~C.}\ \bibnamefont {Thom}}, \bibinfo
  {author} {\bibfnamefont {E.}~\bibnamefont {Tolkacheva}}, \bibinfo {author}
  {\bibfnamefont {C.~J.~S.}\ \bibnamefont {Truncik}}, \bibinfo {author}
  {\bibfnamefont {S.}~\bibnamefont {Uchaikin}}, \bibinfo {author}
  {\bibfnamefont {J.}~\bibnamefont {Wang}}, \bibinfo {author} {\bibfnamefont
  {B.}~\bibnamefont {Wilson}}, \ and\ \bibinfo {author} {\bibfnamefont
  {G.}~\bibnamefont {Rose}},\ }\href {\doibase 10.1038/nature10012} {\bibfield
  {journal} {\bibinfo  {journal} {Nature}\ }\textbf {\bibinfo {volume} {473}},\
  \bibinfo {pages} {194} (\bibinfo {year} {2011})}\BibitemShut {NoStop}%
\bibitem [{\citenamefont {Nelson}\ \emph {et~al.}(2022)\citenamefont {Nelson},
  \citenamefont {Vuffray}, \citenamefont {Lokhov}, \citenamefont {Albash},\
  and\ \citenamefont {Coffrin}}]{Nelson2022-hb}%
  \BibitemOpen
  \bibfield  {author} {\bibinfo {author} {\bibfnamefont {J.}~\bibnamefont
  {Nelson}}, \bibinfo {author} {\bibfnamefont {M.}~\bibnamefont {Vuffray}},
  \bibinfo {author} {\bibfnamefont {A.~Y.}\ \bibnamefont {Lokhov}}, \bibinfo
  {author} {\bibfnamefont {T.}~\bibnamefont {Albash}}, \ and\ \bibinfo {author}
  {\bibfnamefont {C.}~\bibnamefont {Coffrin}},\ }\href@noop {} {\bibfield
  {journal} {\bibinfo  {journal} {Phys. Rev. Applied}\ }\textbf {\bibinfo
  {volume} {17}},\ \bibinfo {pages} {044046} (\bibinfo {year}
  {2022})}\BibitemShut {NoStop}%
\bibitem [{\citenamefont {Raymond}\ \emph {et~al.}(2016)\citenamefont
  {Raymond}, \citenamefont {Yarkoni},\ and\ \citenamefont
  {Andriyash}}]{Raymond2016-je}%
  \BibitemOpen
  \bibfield  {author} {\bibinfo {author} {\bibfnamefont {J.}~\bibnamefont
  {Raymond}}, \bibinfo {author} {\bibfnamefont {S.}~\bibnamefont {Yarkoni}}, \
  and\ \bibinfo {author} {\bibfnamefont {E.}~\bibnamefont {Andriyash}},\
  }\href@noop {} {\bibfield  {journal} {\bibinfo  {journal} {Frontiers in ICT}\
  }\textbf {\bibinfo {volume} {3}} (\bibinfo {year} {2016})}\BibitemShut
  {NoStop}%
\bibitem [{\citenamefont {Marshall}\ \emph {et~al.}(2019)\citenamefont
  {Marshall}, \citenamefont {Venturelli}, \citenamefont {Hen},\ and\
  \citenamefont {Rieffel}}]{Marshall2019-fp}%
  \BibitemOpen
  \bibfield  {author} {\bibinfo {author} {\bibfnamefont {J.}~\bibnamefont
  {Marshall}}, \bibinfo {author} {\bibfnamefont {D.}~\bibnamefont
  {Venturelli}}, \bibinfo {author} {\bibfnamefont {I.}~\bibnamefont {Hen}}, \
  and\ \bibinfo {author} {\bibfnamefont {E.~G.}\ \bibnamefont {Rieffel}},\
  }\href@noop {} {\bibfield  {journal} {\bibinfo  {journal} {Phys. Rev. Appl.}\
  }\textbf {\bibinfo {volume} {11}},\ \bibinfo {pages} {044083} (\bibinfo
  {year} {2019})}\BibitemShut {NoStop}%
\bibitem [{\citenamefont {Marshall}\ \emph {et~al.}(2017)\citenamefont
  {Marshall}, \citenamefont {Rieffel},\ and\ \citenamefont
  {Hen}}]{Marshall2017-rv}%
  \BibitemOpen
  \bibfield  {author} {\bibinfo {author} {\bibfnamefont {J.}~\bibnamefont
  {Marshall}}, \bibinfo {author} {\bibfnamefont {E.~G.}\ \bibnamefont
  {Rieffel}}, \ and\ \bibinfo {author} {\bibfnamefont {I.}~\bibnamefont
  {Hen}},\ }\href@noop {} {\bibfield  {journal} {\bibinfo  {journal} {Phys.
  Rev. Appl.}\ }\textbf {\bibinfo {volume} {8}},\ \bibinfo {pages} {064025}
  (\bibinfo {year} {2017})}\BibitemShut {NoStop}%
\bibitem [{\citenamefont {Sandt}\ and\ \citenamefont
  {Spatschek}(2023)}]{Sandt2023-vu}%
  \BibitemOpen
  \bibfield  {author} {\bibinfo {author} {\bibfnamefont {R.}~\bibnamefont
  {Sandt}}\ and\ \bibinfo {author} {\bibfnamefont {R.}~\bibnamefont
  {Spatschek}},\ }\href@noop {} {\bibfield  {journal} {\bibinfo  {journal}
  {Sci. Rep.}\ }\textbf {\bibinfo {volume} {13}},\ \bibinfo {pages} {6754}
  (\bibinfo {year} {2023})}\BibitemShut {NoStop}%
\bibitem [{\citenamefont {Vuffray}\ \emph {et~al.}(2022)\citenamefont
  {Vuffray}, \citenamefont {Coffrin}, \citenamefont {Kharkov},\ and\
  \citenamefont {Lokhov}}]{Vuffray2022-gr}%
  \BibitemOpen
  \bibfield  {author} {\bibinfo {author} {\bibfnamefont {M.}~\bibnamefont
  {Vuffray}}, \bibinfo {author} {\bibfnamefont {C.}~\bibnamefont {Coffrin}},
  \bibinfo {author} {\bibfnamefont {Y.~A.}\ \bibnamefont {Kharkov}}, \ and\
  \bibinfo {author} {\bibfnamefont {A.~Y.}\ \bibnamefont {Lokhov}},\
  }\href@noop {} {\bibfield  {journal} {\bibinfo  {journal} {PRX Quantum}\
  }\textbf {\bibinfo {volume} {3}},\ \bibinfo {pages} {020317} (\bibinfo {year}
  {2022})}\BibitemShut {NoStop}%
\bibitem [{\citenamefont {Matsuda}\ \emph {et~al.}(2009)\citenamefont
  {Matsuda}, \citenamefont {Nishimori},\ and\ \citenamefont
  {Katzgraber}}]{Matsuda2009-cs}%
  \BibitemOpen
  \bibfield  {author} {\bibinfo {author} {\bibfnamefont {Y.}~\bibnamefont
  {Matsuda}}, \bibinfo {author} {\bibfnamefont {H.}~\bibnamefont {Nishimori}},
  \ and\ \bibinfo {author} {\bibfnamefont {H.~G.}\ \bibnamefont {Katzgraber}},\
  }\href@noop {} {\bibfield  {journal} {\bibinfo  {journal} {New J. Phys.}\
  }\textbf {\bibinfo {volume} {11}},\ \bibinfo {pages} {073021} (\bibinfo
  {year} {2009})}\BibitemShut {NoStop}%
\bibitem [{\citenamefont {Liu}\ and\ \citenamefont {Lee}(2017)}]{Liu2017-nn}%
  \BibitemOpen
  \bibfield  {author} {\bibinfo {author} {\bibfnamefont {Q.}~\bibnamefont
  {Liu}}\ and\ \bibinfo {author} {\bibfnamefont {J.}~\bibnamefont {Lee}},\ }in\
  \href@noop {} {\emph {\bibinfo {booktitle} {Proceedings of the 20th
  International Conference on Artificial Intelligence and Statistics}}},\
  \bibinfo {series} {Proceedings of Machine Learning Research}, Vol.~\bibinfo
  {volume} {54}\ (\bibinfo  {publisher} {PMLR},\ \bibinfo {year} {2017})\ pp.\
  \bibinfo {pages} {952--961}\BibitemShut {NoStop}%
\bibitem [{\citenamefont {Hodgkinson}\ \emph {et~al.}(2020)\citenamefont
  {Hodgkinson}, \citenamefont {Salomone},\ and\ \citenamefont
  {Roosta}}]{Hodgkinson2020-cr}%
  \BibitemOpen
  \bibfield  {author} {\bibinfo {author} {\bibfnamefont {L.}~\bibnamefont
  {Hodgkinson}}, \bibinfo {author} {\bibfnamefont {R.}~\bibnamefont
  {Salomone}}, \ and\ \bibinfo {author} {\bibfnamefont {F.}~\bibnamefont
  {Roosta}},\ }\href@noop {} {\bibfield  {journal} {\bibinfo  {journal}
  {arXiv}\ } (\bibinfo {year} {2020})},\ \Eprint
  {http://arxiv.org/abs/2001.09266} {2001.09266 [math.ST]} \BibitemShut
  {NoStop}%
\bibitem [{\citenamefont {Pelofske}\ \emph {et~al.}(2021)\citenamefont
  {Pelofske}, \citenamefont {Golden}, \citenamefont {B{\"a}rtschi},
  \citenamefont {O'Malley},\ and\ \citenamefont
  {Eidenbenz}}]{pelofske2021sampling}%
  \BibitemOpen
  \bibfield  {author} {\bibinfo {author} {\bibfnamefont {E.}~\bibnamefont
  {Pelofske}}, \bibinfo {author} {\bibfnamefont {J.}~\bibnamefont {Golden}},
  \bibinfo {author} {\bibfnamefont {A.}~\bibnamefont {B{\"a}rtschi}}, \bibinfo
  {author} {\bibfnamefont {D.}~\bibnamefont {O'Malley}}, \ and\ \bibinfo
  {author} {\bibfnamefont {S.}~\bibnamefont {Eidenbenz}},\ }in\ \href@noop {}
  {\emph {\bibinfo {booktitle} {2021 IEEE International Conference on Quantum
  Computing and Engineering (QCE)}}}\ (\bibinfo {organization} {IEEE},\
  \bibinfo {year} {2021})\ pp.\ \bibinfo {pages} {207--217}\BibitemShut
  {NoStop}%
\bibitem [{\citenamefont {Yang}\ \emph {et~al.}(2018)\citenamefont {Yang},
  \citenamefont {Liu}, \citenamefont {Rao},\ and\ \citenamefont
  {Neville}}]{Yang2018-pa}%
  \BibitemOpen
  \bibfield  {author} {\bibinfo {author} {\bibfnamefont {J.}~\bibnamefont
  {Yang}}, \bibinfo {author} {\bibfnamefont {Q.}~\bibnamefont {Liu}}, \bibinfo
  {author} {\bibfnamefont {V.}~\bibnamefont {Rao}}, \ and\ \bibinfo {author}
  {\bibfnamefont {J.}~\bibnamefont {Neville}},\ }in\ \href@noop {} {\emph
  {\bibinfo {booktitle} {Proceedings of the 35th International Conference on
  Machine Learning}}},\ \bibinfo {series} {Proceedings of Machine Learning
  Research}, Vol.~\bibinfo {volume} {80}\ (\bibinfo  {publisher} {PMLR},\
  \bibinfo {year} {2018})\ pp.\ \bibinfo {pages} {5561--5570}\BibitemShut
  {NoStop}%
\bibitem [{\citenamefont {Rahimi}\ and\ \citenamefont
  {Recht}(2007)}]{Rahimi2007-ay}%
  \BibitemOpen
  \bibfield  {author} {\bibinfo {author} {\bibfnamefont {A.}~\bibnamefont
  {Rahimi}}\ and\ \bibinfo {author} {\bibfnamefont {B.}~\bibnamefont {Recht}},\
  }\href@noop {} {\bibfield  {journal} {\bibinfo  {journal} {Adv. Neural Inf.
  Process. Syst.}\ }\textbf {\bibinfo {volume} {20}} (\bibinfo {year}
  {2007})}\BibitemShut {NoStop}%
\bibitem [{\citenamefont {Kivinen}\ and\ \citenamefont
  {Warmuth}(1997)}]{Kivinen1997-va}%
  \BibitemOpen
  \bibfield  {author} {\bibinfo {author} {\bibfnamefont {J.}~\bibnamefont
  {Kivinen}}\ and\ \bibinfo {author} {\bibfnamefont {M.~K.}\ \bibnamefont
  {Warmuth}},\ }\href@noop {} {\bibfield  {journal} {\bibinfo  {journal}
  {Inform. and Comput.}\ }\textbf {\bibinfo {volume} {132}},\ \bibinfo {pages}
  {1} (\bibinfo {year} {1997})}\BibitemShut {NoStop}%
\bibitem [{min()}]{minorminer}%
  \BibitemOpen
  \href@noop {} {\enquote {\bibinfo {title} {Dwavesystems/minorminer},}\
  }\bibinfo {howpublished} {\url{https://github.com/dwavesystems/minorminer}},\
  \bibinfo {note} {accessed: 2023-08-03}\BibitemShut {NoStop}%
\bibitem [{\citenamefont {Andersen}\ \emph {et~al.}(2013)\citenamefont
  {Andersen}, \citenamefont {Dahl}, \citenamefont {Vandenberghe} \emph
  {et~al.}}]{cvxopt}%
  \BibitemOpen
  \bibfield  {author} {\bibinfo {author} {\bibfnamefont {M.~S.}\ \bibnamefont
  {Andersen}}, \bibinfo {author} {\bibfnamefont {J.}~\bibnamefont {Dahl}},
  \bibinfo {author} {\bibfnamefont {L.}~\bibnamefont {Vandenberghe}},  \emph
  {et~al.},\ }\href@noop {} {\bibfield  {journal} {\bibinfo  {journal}
  {Available at cvxopt.org}\ }\textbf {\bibinfo {volume} {54}} (\bibinfo {year}
  {2013})}\BibitemShut {NoStop}%
\bibitem [{\citenamefont {Aoki}\ \emph {et~al.}(2012)\citenamefont {Aoki},
  \citenamefont {Hara},\ and\ \citenamefont {Takemura}}]{aoki2012markov}%
  \BibitemOpen
  \bibfield  {author} {\bibinfo {author} {\bibfnamefont {S.}~\bibnamefont
  {Aoki}}, \bibinfo {author} {\bibfnamefont {H.}~\bibnamefont {Hara}}, \ and\
  \bibinfo {author} {\bibfnamefont {A.}~\bibnamefont {Takemura}},\ }\href@noop
  {} {\emph {\bibinfo {title} {Markov bases in algebraic statistics}}}\
  (\bibinfo  {publisher} {Springer Science \& Business Media},\ \bibinfo {year}
  {2012})\BibitemShut {NoStop}%
\bibitem [{\citenamefont {Inagaki}\ \emph {et~al.}(2016)\citenamefont
  {Inagaki}, \citenamefont {Haribara}, \citenamefont {Igarashi}, \citenamefont
  {Sonobe}, \citenamefont {Tamate}, \citenamefont {Honjo}, \citenamefont
  {Marandi}, \citenamefont {McMahon}, \citenamefont {Umeki}, \citenamefont
  {Enbutsu}, \citenamefont {Tadanaga}, \citenamefont {Takenouchi},
  \citenamefont {Aihara}, \citenamefont {Kawarabayashi}, \citenamefont {Inoue},
  \citenamefont {Utsunomiya},\ and\ \citenamefont {Takesue}}]{Inagaki:2016aa}%
  \BibitemOpen
  \bibfield  {author} {\bibinfo {author} {\bibfnamefont {T.}~\bibnamefont
  {Inagaki}}, \bibinfo {author} {\bibfnamefont {Y.}~\bibnamefont {Haribara}},
  \bibinfo {author} {\bibfnamefont {K.}~\bibnamefont {Igarashi}}, \bibinfo
  {author} {\bibfnamefont {T.}~\bibnamefont {Sonobe}}, \bibinfo {author}
  {\bibfnamefont {S.}~\bibnamefont {Tamate}}, \bibinfo {author} {\bibfnamefont
  {T.}~\bibnamefont {Honjo}}, \bibinfo {author} {\bibfnamefont
  {A.}~\bibnamefont {Marandi}}, \bibinfo {author} {\bibfnamefont {P.~L.}\
  \bibnamefont {McMahon}}, \bibinfo {author} {\bibfnamefont {T.}~\bibnamefont
  {Umeki}}, \bibinfo {author} {\bibfnamefont {K.}~\bibnamefont {Enbutsu}},
  \bibinfo {author} {\bibfnamefont {O.}~\bibnamefont {Tadanaga}}, \bibinfo
  {author} {\bibfnamefont {H.}~\bibnamefont {Takenouchi}}, \bibinfo {author}
  {\bibfnamefont {K.}~\bibnamefont {Aihara}}, \bibinfo {author} {\bibfnamefont
  {K.-I.}\ \bibnamefont {Kawarabayashi}}, \bibinfo {author} {\bibfnamefont
  {K.}~\bibnamefont {Inoue}}, \bibinfo {author} {\bibfnamefont
  {S.}~\bibnamefont {Utsunomiya}}, \ and\ \bibinfo {author} {\bibfnamefont
  {H.}~\bibnamefont {Takesue}},\ }\href {\doibase 10.1126/science.aah4243}
  {\bibfield  {journal} {\bibinfo  {journal} {Science}\ }\textbf {\bibinfo
  {volume} {354}},\ \bibinfo {pages} {603} (\bibinfo {year}
  {2016})}\BibitemShut {NoStop}%
\bibitem [{\citenamefont {Mao}\ \emph {et~al.}(2023)\citenamefont {Mao},
  \citenamefont {Matsuda}, \citenamefont {Tamura},\ and\ \citenamefont
  {Tsuda}}]{mao2023chemical}%
  \BibitemOpen
  \bibfield  {author} {\bibinfo {author} {\bibfnamefont {Z.}~\bibnamefont
  {Mao}}, \bibinfo {author} {\bibfnamefont {Y.}~\bibnamefont {Matsuda}},
  \bibinfo {author} {\bibfnamefont {R.}~\bibnamefont {Tamura}}, \ and\ \bibinfo
  {author} {\bibfnamefont {K.}~\bibnamefont {Tsuda}},\ }\href@noop {}
  {\bibfield  {journal} {\bibinfo  {journal} {Digital Discovery}\ }\textbf
  {\bibinfo {volume} {2}},\ \bibinfo {pages} {1098} (\bibinfo {year}
  {2023})}\BibitemShut {NoStop}%
\end{thebibliography}

%

\end{document}